\documentclass[conference]{IEEEtran}
\usepackage{amsmath, amssymb, graphicx, xcolor, multirow, multicol, comment, subcaption, algorithm2e, url, float, etoolbox, adjustbox, pgf, soul, geometry, colortbl, booktabs}
\title{\textit{Strong Alone, Stronger Together}: Synergizing Modality-Binding Foundation Models with Optimal Transport for Non-Verbal Emotion Recognition}

\author{
    \IEEEauthorblockN{
        Orchid Chetia Phukan\IEEEauthorrefmark{1},
        Mohd Mujtaba Akhtar\IEEEauthorrefmark{1}\IEEEauthorrefmark{3}\thanks{*authors contributed equally},
      Girish\IEEEauthorrefmark{1}\IEEEauthorrefmark{2}\IEEEauthorrefmark{3}, 
        Swarup Ranjan Behera\IEEEauthorrefmark{4}, \\
        Sishir Kalita\IEEEauthorrefmark{5},
        Arun Balaji Buduru\IEEEauthorrefmark{1},
        Rajesh Sharma\IEEEauthorrefmark{1}\IEEEauthorrefmark{6}
        S.R Mahadeva Prasanna\IEEEauthorrefmark{7}\IEEEauthorrefmark{8}
    }
    \IEEEauthorblockA{
        \IEEEauthorrefmark{1}\textit{IIIT-Delhi, India}, 
        \IEEEauthorrefmark{2}\textit{UPES, India},
        \IEEEauthorrefmark{4}\textit{Reliance Jio AICoE, India}
        \IEEEauthorrefmark{5}\textit{Armsoftech.air, India}\\
        \IEEEauthorrefmark{6}\textit{University of Tartu, Estonia}, 
        \IEEEauthorrefmark{7}\textit{IIT-Dharwad, India},
        \IEEEauthorrefmark{8}\textit{IIIT-Dharwad, India}\\
        \IEEEauthorrefmark{3}Contributed equally \\
        \texttt{orchidp@iiitd.ac.in}
    }
}

\begin{document}

\maketitle

\begin{abstract}
\noindent In this study, we investigate multimodal foundation models (MFMs) for emotion recognition from non-verbal sounds. We hypothesize that MFMs, with their joint pre-training across multiple modalities, will be more effective in non-verbal sounds emotion recognition (NVER) by better interpreting and differentiating subtle emotional cues that may be ambiguous in audio-only foundation models (AFMs). To validate our hypothesis, we extract representations from state-of-the-art (SOTA) MFMs and AFMs and evaluated them on benchmark NVER datasets. We also investigate the potential of combining selected foundation model representations to enhance NVER further inspired by research in speech recognition and audio deepfake detection. To achieve this, we propose a framework called \textbf{MATA} (Intra-\textbf{M}odality \textbf{A}lignment through \textbf{T}ransport \textbf{A}ttention). Through \textbf{MATA} coupled with the combination of MFMs: LanguageBind and ImageBind, we report the topmost performance with accuracies of 76.47\%, 77.40\%, 75.12\% and F1-scores of 70.35\%, 76.19\%, 74.63\% for ASVP-ESD, JNV, and VIVAE datasets against individual FMs and baseline fusion techniques and report SOTA on the benchmark datasets.

\end{abstract}

\section{Introduction}

Emotion recognition plays a critical role in understanding human behavior, affecting decision-making, interpersonal relationships, and well-being. While emotions can be identified through multiple channels - such as facial expressions, physiological signals, and vocal cues - non-verbal sounds offer a unique and often underexplored perspective. Non-verbal vocalizations, including laughter, cries, and sighs, convey a broad spectrum of emotions that enhance communication in daily life. Recognizing emotions from these non-verbal vocal cues has applications in diverse areas, such as healthcare, human-computer interaction, customer service, and security. In this study, we focus specifically on non-verbal emotion recognition (NVER). 

However, recent research in emotion recognition has largely centered around verbal speech, employing both handcrafted spectral features \cite{kishore2013emotion} and more recently, audio foundation models (AFMs) \cite{chen2023exploring}. AFMs, such as WavLM \cite{diatlova24_odyssey}, wav2vec2 \cite{pepino21_interspeech}, and HuBERT \cite{morais2022speech}, have shown considerable promise in capturing emotional cues in speech. These foundation models (FMs) are typically fine-tuned or used as feature extractors for downstream emotion recognition tasks. While significant progress has been made, 
non-verbal vocalizations remain underrepresented in the field except a few notable ones \cite{hsu2021speech, tzirakis2023large, xin2024jvnv}. Furthermore, multimodal foundation models (MFMs) remain largely unexplored for NVER despite their potential for more nuanced emotional interpretation. 

In this paper, we aim to address this gap by exploring the use of MFMs for NVER. \textit{We hypothesize that MFMs, are better equipped for NVER due to their multimodal pre-training that enhances their contextual understanding, enabling the model to better interpret and differentiate subtle emotional cues in non-verbal sounds that may be ambiguous in AFMs}. To test this hypothesis, we conduct a comparative study of state-of-the-art (SOTA) MFMs (LanguageBind and ImageBind) and AFMs (WavLM, Unispeech-SAT, and Wav2vec2) by extracting their representations and building a simple downstream CNN model on benchmark NVER datasets (ASVP-ESD, JNV, and VIVAE). 


Furthermore, inspired by research in related areas, such as speech recognition \cite{arunkumar22b_interspeech} and audio deepfake detection \cite{chetia-phukan-etal-2024-heterogeneity}, which have demonstrated the effectiveness of combining FMs due to their complementary behavior, we take the first step in NVER toward this direction. For this purpose, we propose \textbf{MATA} (Intra-\textbf{M}odality \textbf{A}lignment through \textbf{T}ransport \textbf{A}ttention) framework for the effective fusion of FMs.
\textbf{MATA} introduces a novel fusion mechanism leveraging optimal transport to align and integrate representations from FMs. Our study shows that \textbf{MATA} with the fusion of ImageBind and LanguageBind outperform all the individual FMs as well as baseline fusion techniques and leads to SOTA results across NVER benchmarks.

\noindent Our contributions are summarized as follows:

\begin{itemize} \item We conduct the first comprehensive comparative study of SOTA MFMs and AFMs, demonstrating the superior performance of MFMs for NVER, surpassing unimodal AFMs. 

\item We introduce a novel fusion framework, \textbf{MATA}, that effectively combines FMs representations. With \textbf{MATA}, we achieve the highest reported performance across multiple NVER benchmark datasets, outperforming both individual FMs and baseline fusion techniques. \end{itemize}

We will share the models and codes curated as part of this research after the review process. 
\section{Foundation Models}

In this section, we provide an overview of the SOTA MFMs and AFMs considered in our study. These models are selected due to their SOTA performance across various benchmarks in their respective domains.

\subsection{Multimodal Foundation Models}


\noindent \textbf{ImageBind\footnote{\url{https://github.com/facebookresearch/ImageBind/tree/main}} (IB)~\cite{girdhar2023imagebind}} learns from images, audio, text, IMU, depth, and thermal data, aligning other modality representations to image representations. It uses InfoNCE-based optimization and transformer architecture and support zero-shot capability. It associates modality pairs without paired training data and demonstrating strong cross-modal generalization.

\noindent \textbf{LanguageBind\footnote{\url{https://github.com/PKU-YuanGroup/LanguageBind}} (LB)~\cite{zhu2023languagebind}} uses language as the anchor modality due to its rich contextual knowledge. It aligns video, depth, audio, and infrared data to a frozen language encoder through contrastive learning. Pre-trained on the VIDAL-10M dataset, LanguageBind achieves SOTA performance across several benchmarks.

\subsection{Audio Foundation Models}

We select the AFMs that has shown SOTA performance in SUPERB \cite{yang21c_interspeech} and pre-trained on large scale diverse speech data. 

\noindent \textbf{WavLM\footnote{\url{https://huggingface.co/microsoft/wavlm-base}}~\cite{chen2022wavlm}} combines masked speech modeling and denoising during pre-training and uses 94k hours of data from VoxPopuli, LibriLight, and GigaSpeech datasets.

\noindent \textbf{UniSpeech-SAT\footnote{\url{https://huggingface.co/microsoft/unispeech-sat-base}} \cite{chen2022unispeech}} uses contrastive utternace-wise loss, speaker-aware learning for SOTA performance in speech processing and trained on 94k hours of Gigaspeech, Voxpopuli, and LibriVox datasets. 

\noindent \textbf{Wav2vec2\footnote{\url{https://huggingface.co/facebook/wav2vec2-base}} \cite{baevski2020wav2vec}} doesn't shows SOTA performance like WavLM and Unispeech-SAT in SUPERB. However, we use it due to its performance in speech emotion recognition \cite{pepino21_interspeech}. It is trained in a self-supervised fashion that masks speech inputs at the latent level and optimizing via contrastive learning.

We resample all the audios to 16kHz before passing to the MFMs and AFMs. The representations are extracted using average pooling from the last hidden layer of the FMs, resulting in dimensions of 1024 for ImageBind and 768 for LanguageBind, WavLM, UniSpeech-SAT, and Wav2vec2.

\begin{figure}[bt]
    \centering
    \includegraphics[scale=0.16]{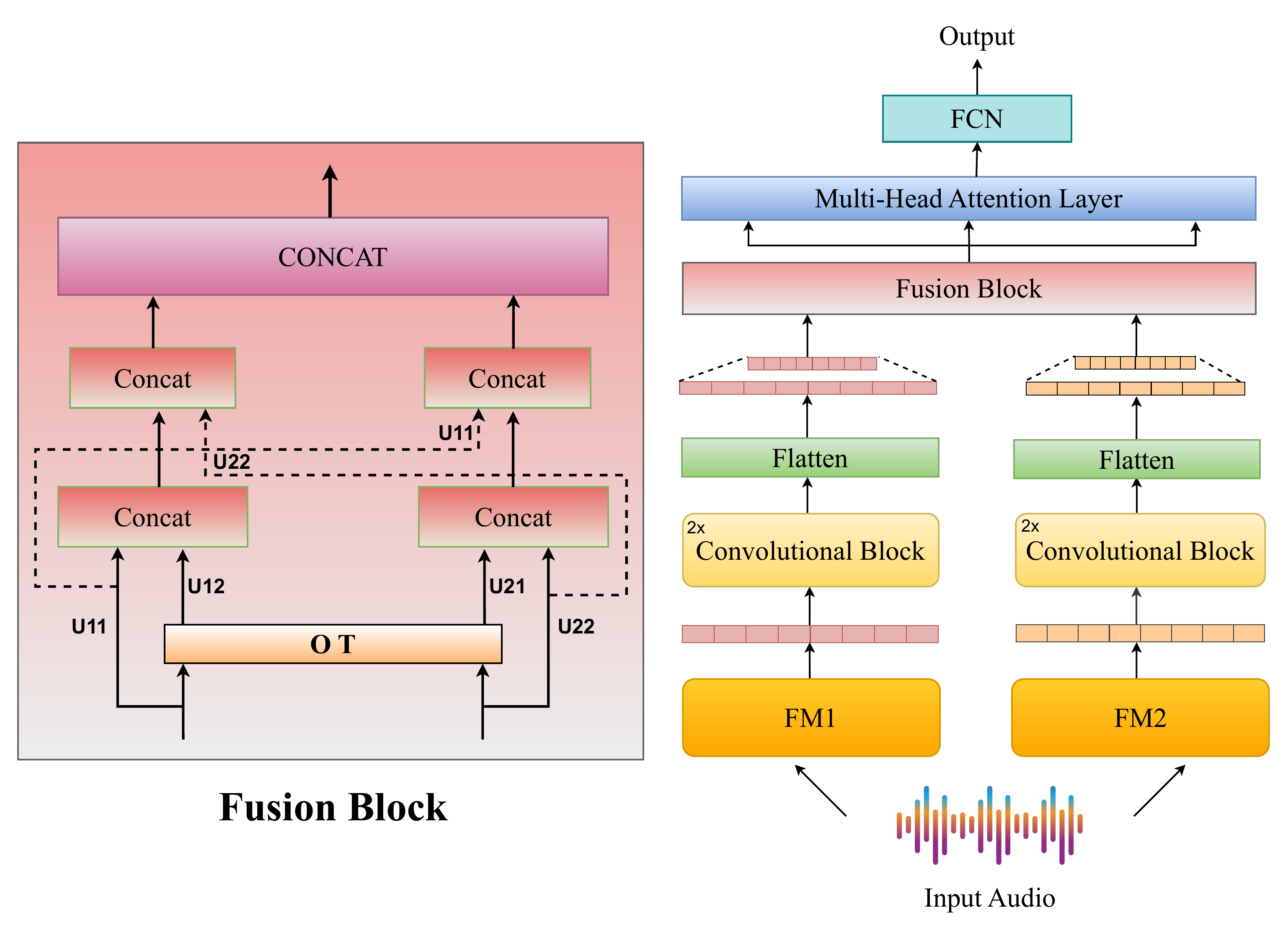}
    \caption{\textbf{MATA} framework: OT and FCN stand for Optimal Transport and Fully Connected Network, respectively. FM1 and FM2 refer to Foundation Model 1 and 2; U11 and U22 represent features from individual FM branches, while U12 and U21 represent features transported from FM2 to the FM1 network and from FM1 to the FM2 network, respectively.}
    \label{fig:work_flow}
\end{figure}

\section{Modeling}

In this section, we discuss the downstream modeling used for individual FMs and the proposed framework, \textbf{MATA} for fusing FMs. 

\subsection{Individual Foundation Models}

The extracted representations from each FM are passed through two convolutional blocks. We experiment with CNN due to its capability shown in related emotion recognition research \cite{chetiaphukan23_interspeech}. Each convolutional block comprises a 1D convolutional layer followed by max-pooling. The first convolutional block uses 64 filters with a kernel size of 3x3, while the second block employs 128 filters with the same size as the first block. The features are then flattened and passed through a dense layer with 128 neurons. Finally, an output layer with softmax activation predicts the emotion classes, matching the number of output neurons to the number of target classes.
The training parameters of the downstream models for different FM representations range from 6.2M to 8.3M.
\begin{table*}[!ht]
\scriptsize
\centering
\caption{Evaluation Scores: Scores are in \% and represent the average of 5 folds. LB, IB, UNI, WA, and WAV2 stands for LanguageBind, ImageBind, Unispeech-SAT, WavLM, and Wav2vec2, respectively. F1-Score is the macro-average F1-Score.}
\label{TAB:1}
\begin{tabular}{l|cc|cc|cc|cc} 
\toprule
\textbf{Features} & \multicolumn{2}{c|}{\textbf{ASVP\_ESD}} & \multicolumn{2}{c|}{\textbf{JNV}} & \multicolumn{2}{c|}{\textbf{VIVAE}} & \multicolumn{2}{c}{\textbf{CREMA-D}} \\ 
\midrule
\textbf{} & \textbf{Accuracy\hspace{0.1cm}} & \textbf{F1-Score\hspace{0.1cm}} & \textbf{Accuracy\hspace{0.1cm}} & \textbf{F1-Score\hspace{0.1cm}} & \textbf{Accuracy\hspace{0.1cm}} & \textbf{F1-Score\hspace{0.1cm}} & \textbf{Accuracy\hspace{0.1cm}} & \textbf{F1-Score} \\ 
\midrule
\multicolumn{9}{c}{\textbf{Individual Representations}} \\ 
\midrule
LB & 75.55 & 67.55 & 73.65 & 72.51 & 69.12 & 68.83 & 63.67 & 63.26 \\ 
IB & 62.03 & 49.11 & 63.10 &  60.97 & 55.30 & 54.63 & 63.67 & 63.68 \\ 
UNI & 49.90 & 35.86 & 57.14 & 53.81 & 36.87 & 35.78 & 63.26 & 63.20 \\ 
WA & 46.98 & 32.77 & 62.07 & 58.33 & 35.94 & 34.39 & 55.88 & 55.92 \\ 
WAV2 & 60.04 & 48.51 & 57.14 & 59.27 & 46.54 & 45.65 & 59.84 & 59.82 \\ 
\midrule
\multicolumn{9}{c}{\textbf{Fusion with Concatenation}} \\ 
\midrule
LB+IB &  \cellcolor{yellow!25}\textbf{76.34} & 64.58 & 72.62 & 72.26 & 67.48 & 67.28 & \cellcolor{green!25}\textbf{71.46} & \cellcolor{green!25}\textbf{71.71} \\ 
LB+UNI & 74.09 & 65.29 & 67.86 & 66.36 & 61.75 & 61.70 & 68.06&  67.70 \\ 
LB+WA & 72.90 & 64.80 & 70.24 & 70.92 & 65.44 & 65.35 & 66.35 & 65.96 \\ 
LB+WAV2 & 73.76 & 62.20 & 70.24 & 69.18 & 60.83 & 60.57 & 69.04 & 68.93 \\ 
IB+UNI & 65.74 & 55.45 & 58.33 & 52.25 & 53.00 & 52.42 & 72.60 & 72.66 \\ 
IB+WA & 66.14 & 56.83 & 58.33 & 52.25 & 58.53 & 57.28 & 69.31 & 69.28 \\ 
IB+WAV2 & 67.20 & 55.53 & 57.14 & 55.19 & 56.22 & 55.73 & 68.57 & 68.67 \\ 
UNI+WA & 53.61 & 38.98 & 44.05 & 39.10 & 44.70 & 43.88 & 66.76 & 66.69 \\ 
UNI+WAV2 & 60.70 & 49.16 & 47.66 & 46.18 & 49.31 & 49.51 & 68.84 & 68.81 \\ 
WA+WAV2 & 59.64 & 45.66 & 51.19 & 41.98 & 48.85 & 46.97 & 68.10 & 68.16 \\ 
\midrule
\multicolumn{9}{c}{\textbf{Fusion with OT}} \\ 
\midrule
LB+IB &  \cellcolor{green!25}\textbf{76.41} &  \cellcolor{yellow!25}\textbf{68.79} & \cellcolor{green!25}\textbf{77.03} & \cellcolor{green!25}\textbf{76.14} & \cellcolor{green!25}\textbf{70.05} & \cellcolor{green!25}\textbf{69.80} & 62.12 & 62.05 \\ 
LB+UNI & 75.61 & 67.52 & 70.24 & 71.90 & 61.29 & 60.33 & 59.44 & 58.84 \\ 
LB+WA & 75.48 & 66.97 & 69.05 & 70.48 & 62.21 & 61.30 & 58.16 & 58.14 \\ 
LB+WAV2 & 75.48 & 67.75 & 67.86 & 66.44 & 66.82 & 66.50 & 58.03 & 57.84 \\ 
IB+UNI & 64.88 & 53.19 & 64.29 & 62.59 & 57.14 & 56.49 & 62.46 & 62.24 \\ 
IB+WA & 64.68 & 54.40 & 59.52 & 59.05 & 57.14 & 56.65 & 59.17 & 59.06 \\ 
IB+WAV2 & 67.00 & 55.41 & 61.90 & 61.21 & 59.91 & 59.30 & 57.76 & 57.62 \\ 
UNI+WA & 55.47 & 45.62 & 59.52 & 56.60 & 41.01 & 39.40 & 60.51 & 60.45 \\ 
UNI+WAV2 & 60.77 & 48.43 & 47.62 & 47.54 & 46.54 & 45.12 & 58.63 & 58.68 \\ 
WA+WAV2 & 60.64 & 51.85 & 60.71 & 60.60 & 49.77 & 49.04 & 58.97 & 59.09 \\ 

\midrule
\multicolumn{9}{c}{\textbf{Fusion with \textbf{MATA}}} \\ 
\midrule
LB+IB & \cellcolor{blue!25}\textbf{76.47} & \cellcolor{blue!25}\textbf{70.35} & \cellcolor{blue!25}\textbf{77.40} & \cellcolor{blue!25}\textbf{76.19} & \cellcolor{blue!25}\textbf{75.12} & \cellcolor{blue!25}\textbf{74.63} & \cellcolor{blue!25}\textbf{72.64} & \cellcolor{blue!25}\textbf{72.62} \\ 
LB+UNI & 75.41 & 66.51 & 71.43 & 72.17 & 65.44 & 64.94 & 69.85 & 69.98 \\ 
LB+WA & 75.75 & \cellcolor{green!25}\textbf{70.25} & 73.81 & \cellcolor{yellow!25}\textbf{73.08} & 69.12 & 68.64 & 66.29 & 66.30 \\ 
LB+WAV2 & 75.81 & 68.49 & \cellcolor{yellow!25}\textbf{75.00} & 72.39 & \cellcolor{yellow!25}\textbf{69.59} & \cellcolor{yellow!25}\textbf{69.15} & 66.55 & 66.66 \\ 
IB+UNI & 67.93 & 60.15 & 61.90 & 62.17 & 60.37 & 59.43 & \cellcolor{yellow!25}\textbf{71.32} & \cellcolor{yellow!25}\textbf{71.49} \\ 
IB+WA & 65.94 & 57.44 & 61.90 & 60.15 & 61.75 & 61.12 & 68.64 & 68.72 \\ 
IB+WAV2 & 68.06 & 61.14 & 64.29 & 64.74 & 60.37 & 60.13 & 68.57 & 68.57 \\ 
UNI+WA & 56.66 & 46.55 & 46.43 & 43.78 & 45.16 & 43.99 & 66.82 & 66.88 \\ 
UNI+WAV2 & 61.43 & 52.89 & 53.57 & 55.02 & 48.85 & 47.58 & 70.99 & 71.06 \\ 
WA+WAV2 & 62.36 & 52.29 & 59.63 & 57.71 & 50.69 & 49.24 & 67.36 & 67.49 \\ 
\bottomrule
\end{tabular}
\end{table*}

\subsection{Modality Alignment through Transport Attention (MATA)}

The architecture of \textbf{MATA} is shown in Figure \ref{fig:work_flow}. For each FM, the extracted representations are passed through two convolutional blocks with the same modeling as used in the individual models above. However, the number of filters used in 1D-CNN in two convolution blocks are 32 and 64. Then, it is flattened, followed by linear projection to 120-dimension. The projection to lower dimensions is due to computational constraints. Then, the features of each network block from individual FMs are passed through the fusion block, which encompasses the optimal transport (OT) distance \( M \) for effective fusion \cite{pramanick2022multimodal} of FMs. \( M \) between the feature matrices, \( x_1 \) and \( x_2 \) from two FMs, computed via normalized Euclidean distance:

\[
M = \frac{\| x_1 - x_2 \|_2}{\max(\| x_1 - x_2 \|_2)}
\]

To align the features, we apply the Sinkhorn algorithm to obtain the optimal transport plan \( \gamma \), where: $\gamma = \text{Sinkhorn}(M) $.
Using \( \gamma \), we transport features between FMs networks, producing \( x_2 \to x_1 \) and \( x_1 \to x_2 \): $
x_2 \to x_1 = \gamma \cdot x_2, \quad x_1 \to x_2 = \gamma^T \cdot x_1$. These transported features are concatenated with the original features from FMs to form the fused representations: $\text{fused}_1 = \text{Concat}(x_2 \to x_1, x_1),
\text{fused}_2 = \text{Concat}(x_1 \to x_2, x_2)$. \par

These fused features are then concatenated with the original features from the opposite FM, as shown in Figure \ref{fig:work_flow}, and the resultant features are finally concatenated and passed to the Multi-Head Attention (MHA) block. The MHA block ensures further better feature interaction due to its self-attention mechanism. The attention output is computed as: 
\[
\text{Attention}(Q, K, V) = \text{softmax}\left(\frac{QK^T}{\sqrt{d_k}}\right) V
\]
where $Q$ and $K$ are the query and key matrices derived from the final concatenated features. $V$ represents the feature vectors that are attended to. Here, we are using multiple attention heads, and the number of heads is 8. The number of training parameters of \textbf{MATA} with different combinations of FMs ranges from 4M to 4.5M. 

\section{Experiments}

\subsection{Benchmark Datasets}
\noindent \textbf{ASVP-ESD \cite{landry2020asvp}:} This dataset includes thousands of high-quality audio recordings labeled with  12 emotions and an additional class breath. The recordings were captured in natural environments with diverse speakers comprising speech and non-speech emotional sounds. We use only the non-speech part in our experiments. The audio samples were gathered from various sources, including films, TV programs, YouTube channels, and other online platforms.

\noindent \textbf{JNV \cite{xin2024jnv}:} It features 420 audio clips from four native Japanese speakers (two male, two female) expressing six emotions: anger, disgust, fear, happiness, sadness, and surprise. Recorded at 48 kHz in an anechoic chamber, the dataset includes both predefined and spontaneous vocalizations. 

\noindent \textbf{VIVAE \cite{holz2022variably}:} It includes 1,085 audio files from eleven speakers expressing three positive (achievement, pleasure, surprise) and three negative emotions (anger, fear, pain) at varying intensities. It was recorded at 44.1 kHz and 16-bit resolution. 

\subsection{Training Details}

We trained our models for 50 epochs with a learning rate of 1e-3 and Adam as the optimizer. We use cross-entropy as the loss function and batch size of 32.  Early stopping and dropout are employed to prevent overfitting.

\begin{figure}[!bt]
    \centering
    \begin{minipage}{0.24\textwidth}
        \centering
        \includegraphics[width=\textwidth]{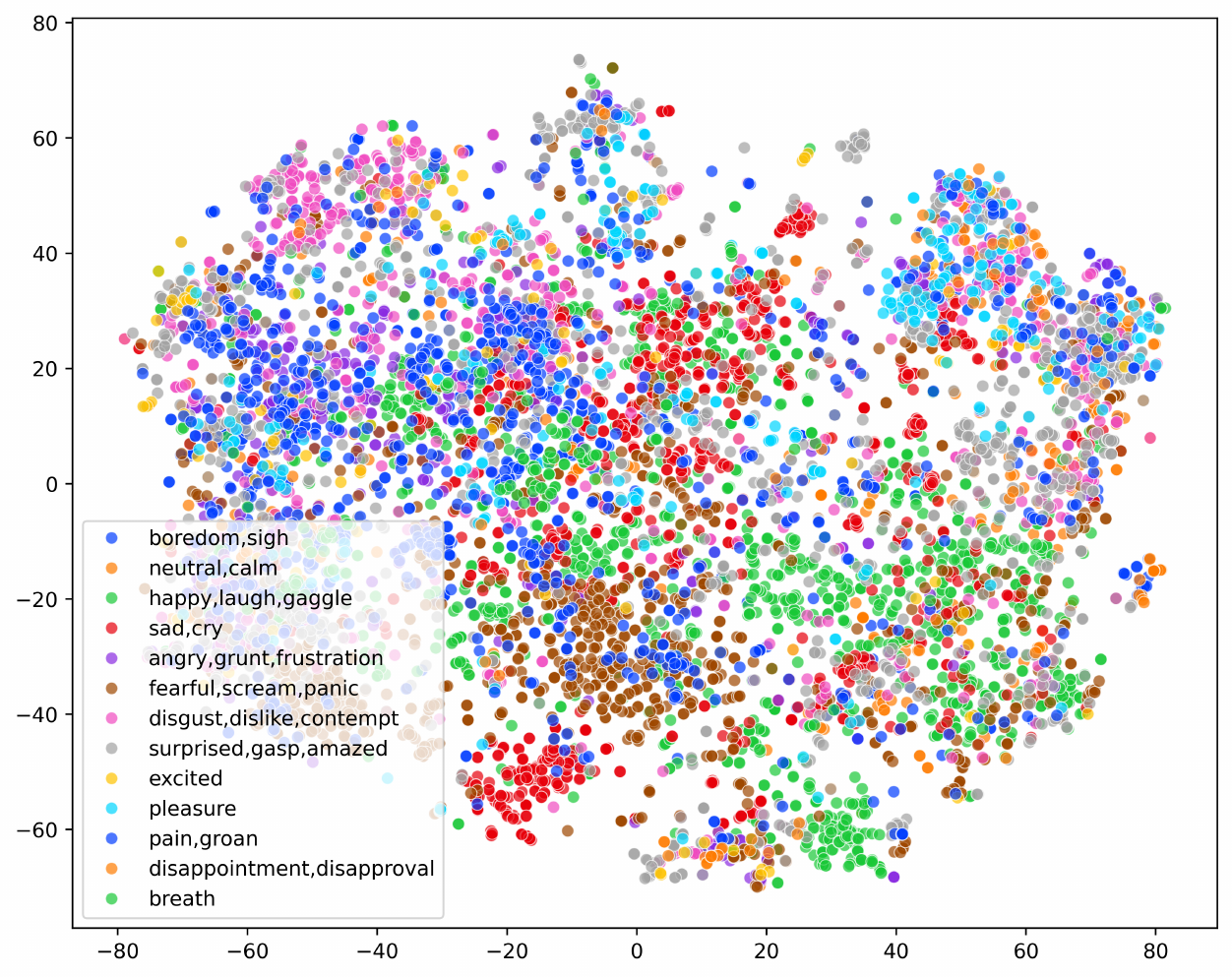} 
        \caption*{(a) ImageBind}
    \end{minipage}\hfill
    \begin{minipage}{0.24\textwidth}
        \centering
        \includegraphics[width=\textwidth]{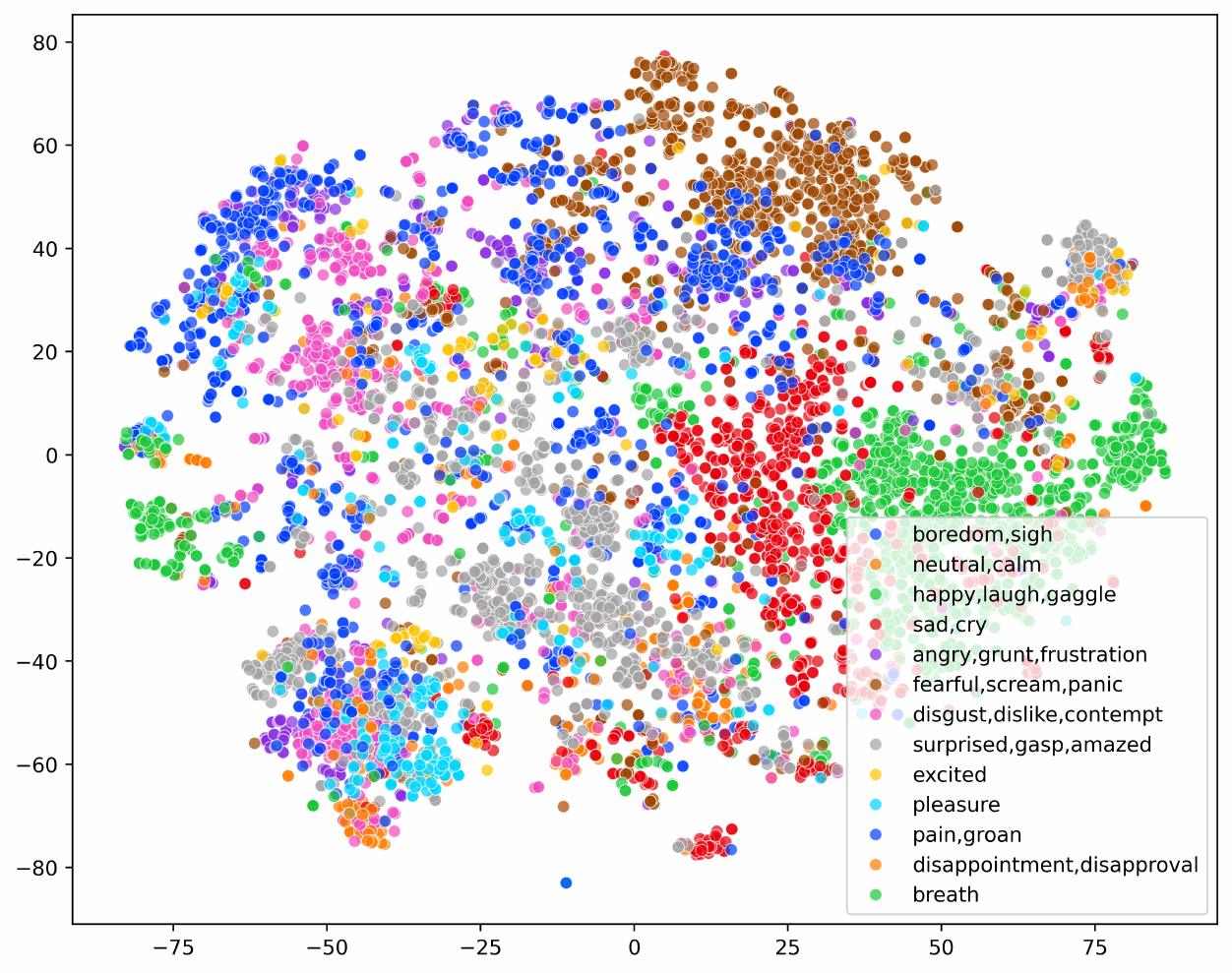} 
        \caption*{(b) LanguageBind}
    \end{minipage}\\[10pt] 
    \begin{minipage}{0.24\textwidth}
        \centering
        \includegraphics[width=\textwidth]{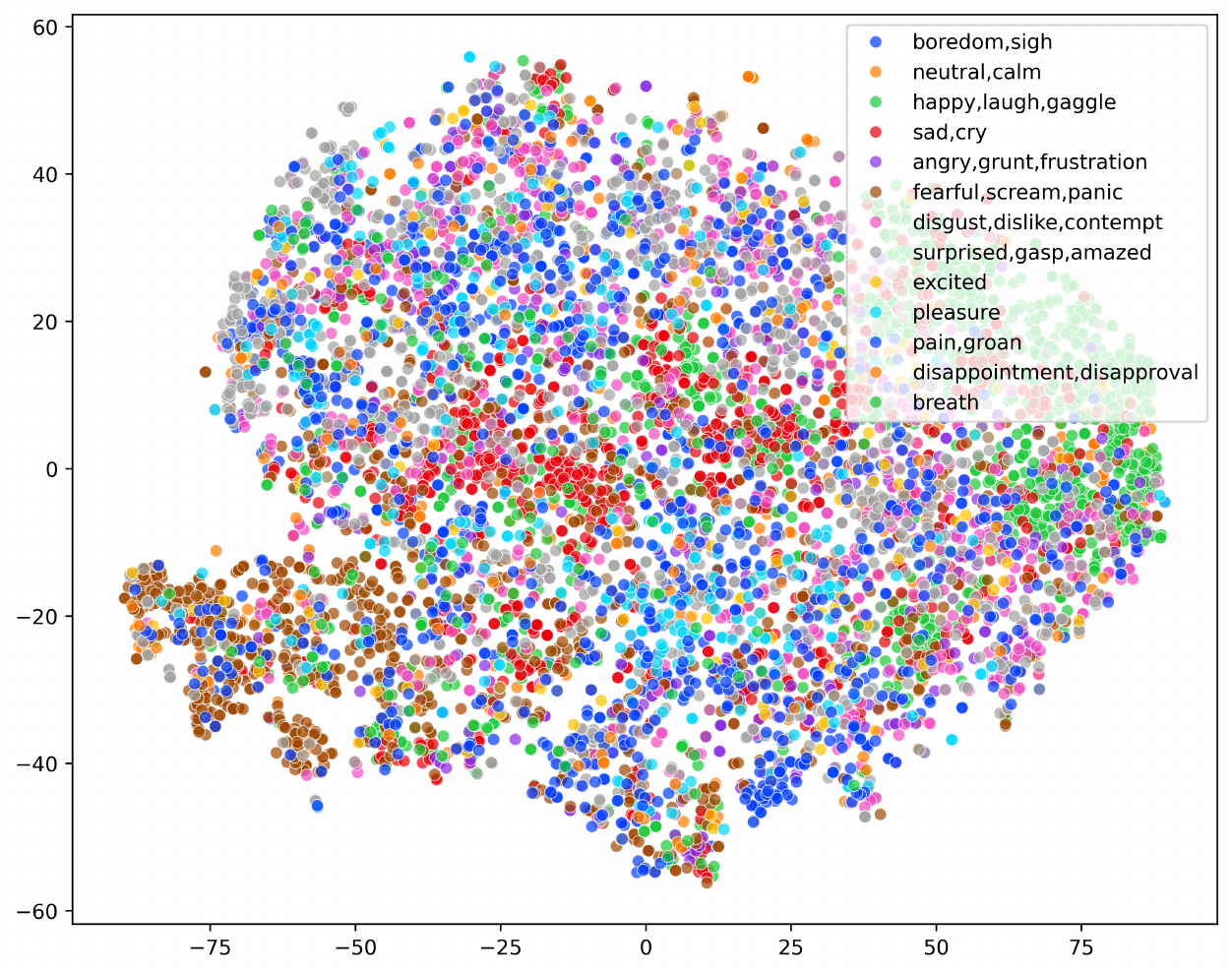} 
        \caption*{(c) WavLM}
    \end{minipage}\hfill
    \begin{minipage}{0.24\textwidth}
        \centering
        \includegraphics[width=\textwidth]{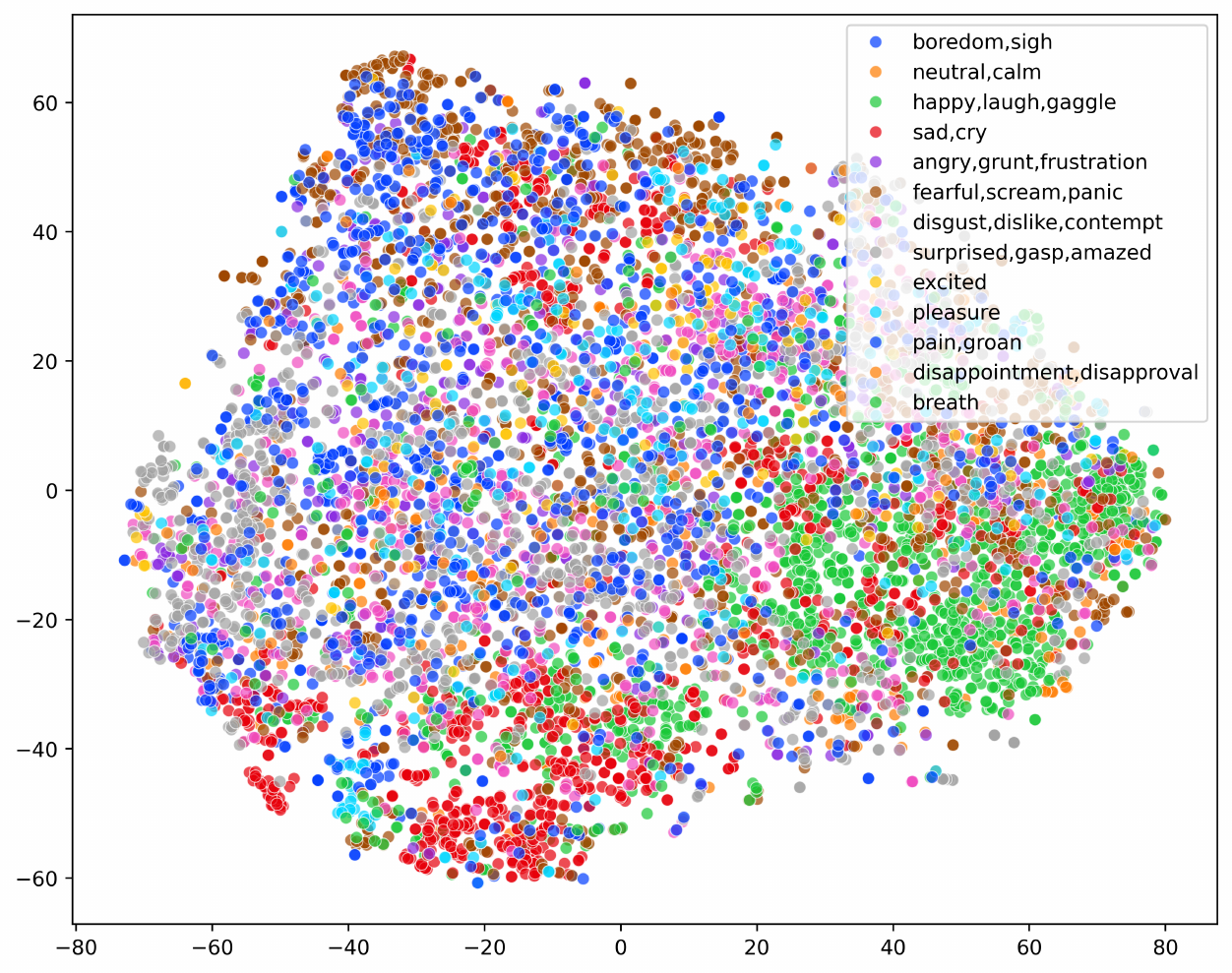} 
        \caption*{(d) Unispeech-SAT}
    \end{minipage}\\[10pt] 
    \caption{t-SNE plots of raw representations from FMs on the ASVP-ESD dataset.} 
    \label{fig:subfigures}
\end{figure}

\begin{figure}[bt]
    \centering
    \begin{subfigure}[b]{0.24\textwidth} 
        \centering
        \includegraphics[width=\linewidth]{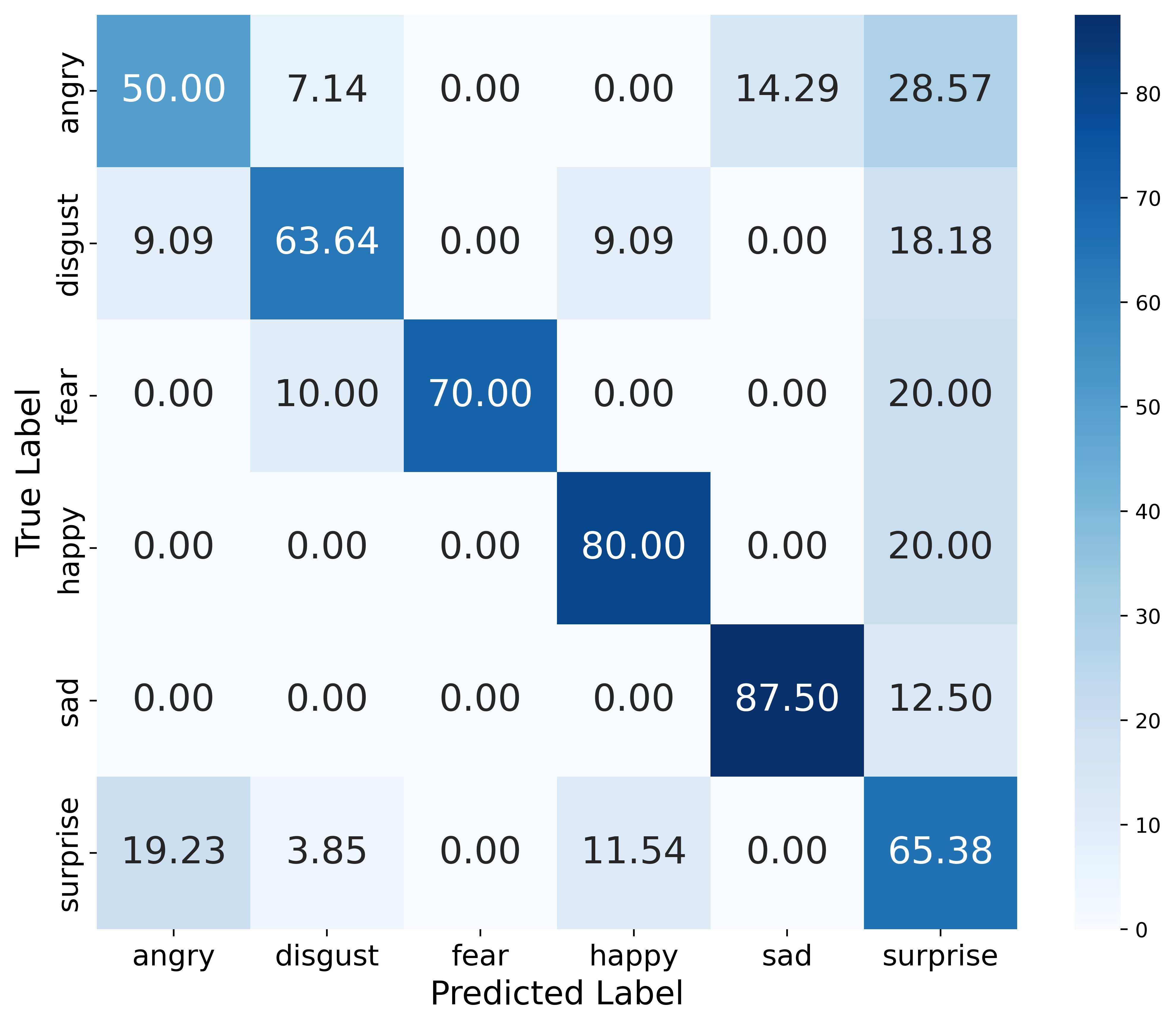}
        \caption{UNI + WA}
    \end{subfigure}
    \hfill
    \begin{subfigure}[b]{0.24\textwidth} 
        \centering
        \includegraphics[width=\linewidth]{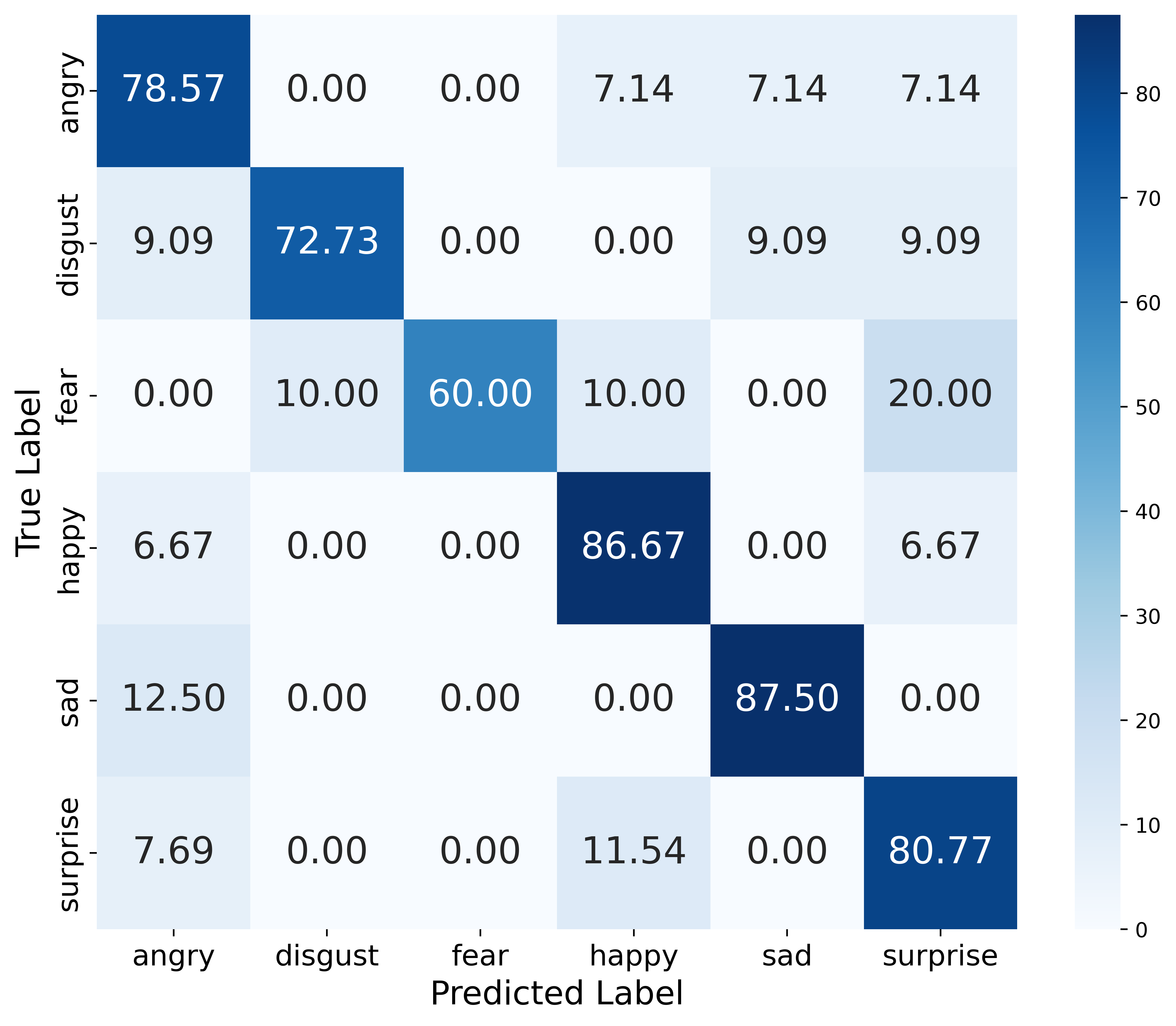}
        \caption{LB + IB}
        \label{fig:histogram}
    \end{subfigure}
    \caption{Confusion Matrix of \textbf{MATA}: UNI, WA, LB, and IB stand for Unispeech-SAT, WavLM, LanguageBind, and ImageBind, respectively.}
    \label{fig:confusion_matrices}
\end{figure}

\subsection{Results and Discussion}

We present the results of the experiments in Table~\ref{TAB:1}. We first evaluated individual FMs. LB achieved the highest performance across all the NVER datasets, with an accuracy of 75.55\% and an F1-score of 67.55\% on ASVP-ESD, 73.65\% accuracy and an F1-score of 72.51\% on JNV, and 70.05\% accuracy and an F1-score of 69.80\% on VIVAE, significantly outperforming all other FMs. 
In summary, the MFMs perform better than the AFMs for NVER, thus proving \
\textit{our hypothesis that MFMs capture complex emotional nuances due to their multimodal pre-training that may be ambiguous to AFMs}. The t-SNE plots of the raw representations from the FMs are shown in Figure \ref{fig:subfigures}. We observe better clusters across emotions for MFMs in comparison to the AFMs. \par

When combining the FMs through \textbf{MATA}, we obtain the topmost performance against all the individual FMs and the baseline concatenation-based fusion technique. In concatenation-based fusion, we use the same architectural components as \textbf{MATA}. This shows the observable complementary behavior of the MFMs as well as the effectiveness of \textbf{MATA} in performing effective fusion of the MFMs. With \textbf{MATA}, we also observe that the fusion of MFMs and AFMs gives comparatively better results than individual FMs as well as the baseline concatenation-based fusion technique. The confusion matrices of \textbf{MATA}, with Unispeech-SAT + WavLM and LanguagebIND + ImageBind are shown in Figure~\ref{fig:confusion_matrices}. We also provide an ablation study of \textbf{MATA} without the MHA block (Table \ref{TAB:1}: Fusion with OT); we observe better results than the individual FMs, comparative results, and sometimes better performance with some pairs of FMs. \par

\noindent \textbf{Additional Experiments:} To show the generalizability of the proposed framework, \textbf{MATA}, we also experimented on a benchmark speech emotion recognition (SER) dataset, CREMA-D \cite{cao2014crema}. It consists of 7,442 clips from 91 actors (48 male, 43 female) expressing six basic emotions: happiness, sadness, anger, fear, disgust, and neutral. Rated by 2,443 participants across audio-only, visual-only, and audio-visual modalities. Due to the diversity of the speakers, CREMA-D serve as essential benchmark for emotion recognition systems. From Table \ref{TAB:1}, we observe that MFMs show better performance than AFMs. However, we achieve the topmost performance with \textbf{MATA} with the combination of LanguageBind and ImageBind representations, thus showing the effectiveness of the proposed framework. 

\vspace{-0.3cm}
\section{Conclusion}
Our study demonstrates the effectiveness of MFMs for NVER. This performance can be attributed to their joint pre-training across multiple modalities that provide better contextual understanding and excel in capturing subtle emotional cues that AFMs may miss. Through extensive evaluation of the benchmark NVER datasets, we confirm the superior performance of MFMs (LanguageBind and ImageBind) in comparison to AFMs such as WavLM, Unispeech-SAT and Wav2vec2. We show more improved performance through the fusion of the FMs by proposing \textbf{MATA} for effective fusion. With \textbf{MATA}, we achieve top performance against all the individual FMs as well as baseline fusion techniques, thus achieving SOTA performance across the NVER benchmark datasets under consideration. Our study provides valuable insights for future research in selecting optimal representations for NVER and usage of MFMs. It also opens pathways for developing effective fusion techniques for the fusion of FMs. 

\bibliographystyle{IEEEtran}
\bibliography{mybib}

\end{document}